\documentclass[aps,pra,showpacs,reprint,superscriptaddress,lengthcheck]{revtex4-1}
\pdfoutput=1
\usepackage{graphicx}
\usepackage{dcolumn}
\usepackage{bm}%
\usepackage{amsmath}
\usepackage{amssymb}
\usepackage{changes}

\usepackage{footnote}

\newcommand{\ko}[2]{\big[ #1, \, #2 \big]}
\newcommand{\ewqm}[1]{\big\langle #1 \big\rangle}
\newcommand{\w}{\omega}

\newcommand{\refk}[1]{(\ref{#1})}

\let\emph\textit

\begin{document}


\title{Time-delayed quantum coherent Pyragas feedback control of photon squeezing in a degenerate parametric oscillator}


\author{Manuel Kraft}
\email{kraft@itp.tu-berlin.de}
\affiliation{Technische Universit\"at Berlin, Institut f\"ur Theoretische Physik, Nichtlineare Optik und Quantenelektronik, Hardenbergstra\ss e 36, 10623 Berlin, Germany}
\author{Sven M. Hein}
\affiliation{Technische Universit\"at Berlin, Institut f\"ur Theoretische Physik, Nichtlineare Optik und Quantenelektronik, Hardenbergstra\ss e 36, 10623 Berlin, Germany}
\author {Judith Lehnert}
\affiliation{Technische Universit\"at Berlin, Institut f\"ur Theoretische Physik, Nichtlineare Dynamik und Kontrolle, Hardenbergstra\ss e 36, 10623 Berlin, Germany}

\author {Eckehard Sch\"oll}
\affiliation{Technische Universit\"at Berlin, Institut f\"ur Theoretische Physik, Nichtlineare Dynamik und Kontrolle, Hardenbergstra\ss e 36, 10623 Berlin, Germany}

\author {Stephen Hughes}
\affiliation{Department of Physics, Engineering Physics and Astronomy, Queen's University, Kingston, Ontario, Canada, K7L 3N6}

\author {Andreas Knorr}
\affiliation{Technische Universit\"at Berlin, Institut f\"ur Theoretische Physik, Nichtlineare Optik und Quantenelektronik, Hardenbergstra\ss e 36, 10623 Berlin, Germany}

\date{\today}

\begin{abstract}

Quantum coherent feedback control is a measurement-free control method fully preserving quantum coherence. In this paper we show how time-delayed quantum coherent feedback can be used to control the degree of squeezing in the output field of a cavity containing a degenerate parametric oscillator. We focus on the specific situation of Pyragas-type feedback control where time-delayed signals are fed back directly into the quantum system. Our results show how time-delayed feedback can enhance or decrease the degree of squeezing as a function of time delay and feedback strength.

\end{abstract}


\maketitle

\section{Introduction}

Quantum coherent feedback control \cite{lloyd2000} has become an increasing field of research in the past years \cite{serafini2012feedback}. In contrast to measurement based quantum control schemes, coherent feedback control does not require quantum state projection by a measurement, and therefore fully preserves the quantum coherence. This offers new interesting possibilities for the control of quantum systems. In particular, coherent time-delayed feedback has been proposed as non-invasive Pyragas-type \cite{pyragas1992, Hoevel2005, Schoell2008book, Grebogi2010} control scenarios \cite{carmele2013, grimsmo2014}, entanglement control in a quantum node network \cite{Hein2015, Zoller2015}, enhancement of atomic lifetimes by shaping the vacuum \cite{hoi2015} or controlling atoms in cavities in front of a mirror \cite{dornerzoller2002, grimsmo2014}.

Squeezed states of light \cite{Carmichael2008, CarmichaelSPIE} have found important application in a broad area of quantum optics and quantum information processing, providing for example an entanglement resource for many quantum information protocols such as quantum key distribution schemes \cite{Ralph2003}, or a signature of quantum synchronization and quantum chimeras \cite{Bastidas2015}.
   
Much work has been done in the control of squeezing of a degenerate parametric oscillator (DPO) in quantum optics in recent years. An early attempt \cite{wiseman1995feedback, wiseman} used a measurement-based feedback scheme to enhance the squeezing of a cavity mode. Recent theoretical research in coherent feedback control introduces a feedback loop through a beam splitter \cite{gough2009, gough2010}. Experimental verification in the instantaneous limit \cite{Crisafulli2013} and time-delayed feedback \cite{Iida2012} has shown the possibility, but also limitations of coherent control of squeezing. 
\textit{Most of the recent work is done in the instantaneous feedback limit.  However time delays in quantum systems are often unavoidable.} 

\textit{In this paper, we study the squeezing spectrum of the output field of a cavity containing a degenerate parametric oscillator controlled by quantum coherent time-delayed feedback.} We focus on the specific situation of Pyragas control, where the difference of instantaneous and time-delayed signal acts as a control force and is fed back into the quantum system \cite{carmele2013, grimsmo2015}. Pyragas control is usually used in classical systems to stabilize unstable periodic orbits. An advantage is that in the case of stabilization the control force vanishes making the control scheme non-invasive. Another characteristic is that no detailed model of the system dynamics has to be known, since it is sufficient to measure an output signal to construct the difference between instantaneous and time-delayed signal.

For our purpose we introduce a quantum mechanical description of time-delayed quantum coherent feedback in the framework of the input-output formalism introduced by Gardiner and Collett \cite{gardinercollett1985inputoutput}. The proposed scheme can be realized in a number of ways, e.g., by a mirror or a nanophotonic arrangement using photonic crystal waveguides \cite{yao2009, Hughes2007, Hughes2007singleauthor, shen2009}.  
  
The paper is organized as follows: First, we will introduce our system and feedback scheme and derive the equation of motion for the internal dynamics and the output fields using input-output theory presented in the appendix. Second, to achieve realistic control conditions we will make a stability analysis to obtain the asymptotic behavior of the controlled system. Third, we calculate the squeezing spectrum of the output fields and discuss how the feedback scheme influences the squeezing performance. In the appendix we will show thoroughly that the dynamics of a system coherently interacting with itself through a feedback-loop can be represented as a cascaded system where previous versions of the system drive the present system leading to time-delayed quantum Langevin equations in the Heisenberg picture~\cite{gardinercollett1985inputoutput}. A key advantage of this approach is that it allows physical insight how feedback coherently returns the outcoupled information back into the system.

\section{Degenerate parametric oscillators} 
\label{sec: degenerate parametric oscilators}

The degenerate parametric oscillator (DPO) is a well-known non-linear optical device which allows for squeezing of externally applied light-fields. We study the effect of feedback on the steady-state squeezing spectrum of the output fields quadrature phases.     

\begin{figure}
 \includegraphics[width=\linewidth]{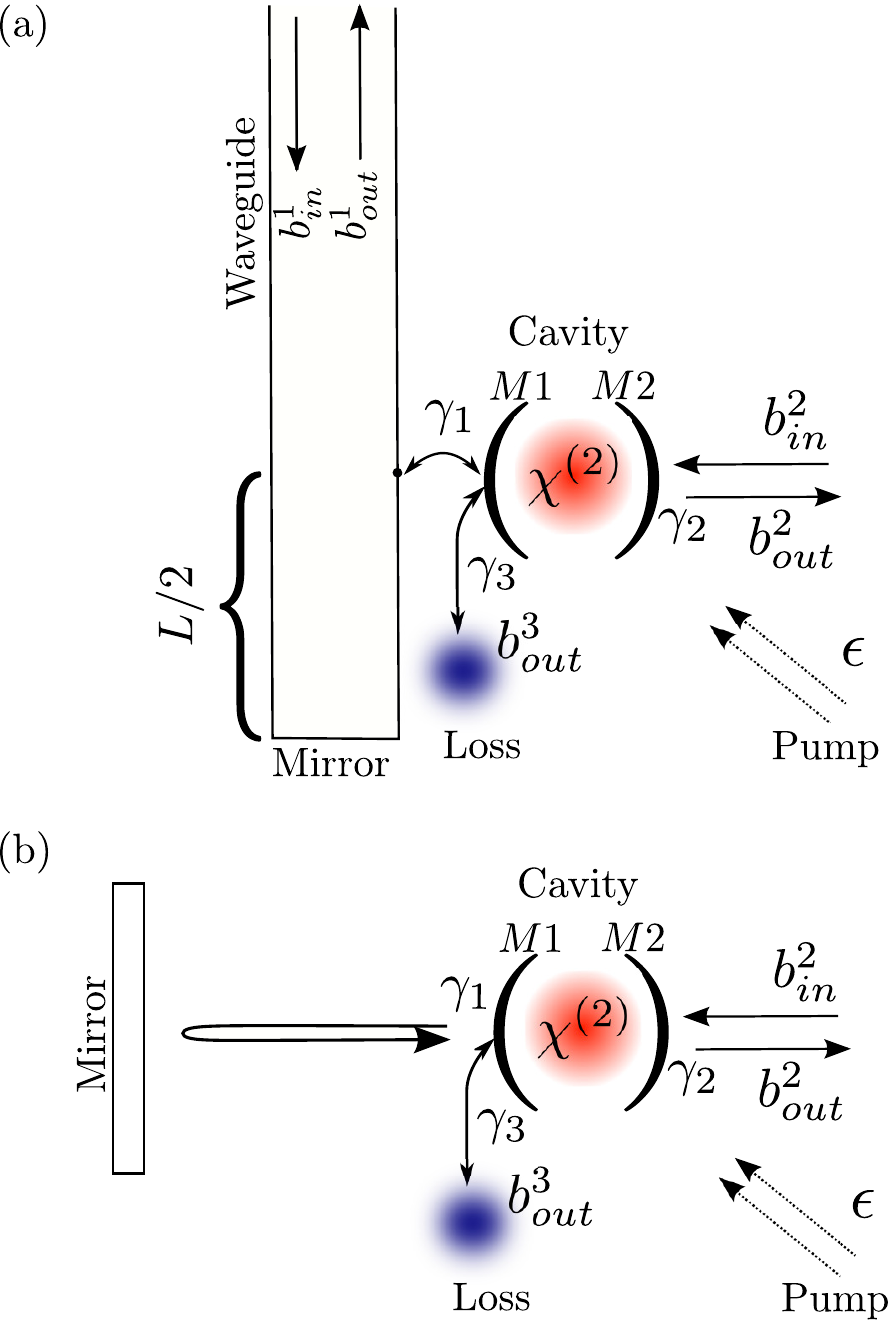}
 \caption{\label{squeezingOPO} (Color online) Schematic of our physical setup and model. A degenerate parametric oscillator (DPO) is embedded in a cavity and driven by a laser with pumping strength $|\epsilon|$. The cavity mirror $M_2$ with loss rate $\gamma_2$ is coupled to an external reservoir denoted by the input $b^2_{in}$ and output $b^2_{out}$. (a) The mirror $M_1$ is coupled to a photonic waveguide terminated by a mirror at a distance $L/2$ from the coupling point. The waveguide mirror will introduce a phase-shift $\phi=\pi$ for the reflected field and time-delayed feedback into the system. For the case of imperfect coupling to the waveguide, we additionally implement another loss-reservoir, graphically denoted by the blue cloud and coupling strength $\gamma_3$. (b) Instead of a waveguide we let the outgoing radiation from $M_1$ directly emit onto an external mirror.}   
\end{figure}

The DPO, a second-order nonlinear crystal (susceptibility $\chi^{(2)}$), is embedded in a two-sided cavity and externally pumped, see Fig. \ref{squeezingOPO}. The cavity-mirrors $M1$ and $M2$ are assumed to be lossy with different photon decay loss rates, but assumed to be transparent at the pump field frequency $\omega_p$ \citep{gardinerzoller2010book}. The output field of $M1$ is redirected back to $M1$, becoming a new input. Such a setup will introduce time-delay in the equation of motion of the system operator (see below). It can be realized in different ways, with two examples shown in Fig. \ref{squeezingOPO}(a,b):

\paragraph*{Case (a), Feedback from a semi-infinite waveguide (Fig. \ref{squeezingOPO}(a))}

 The first possibility is to introduce feedback to the DPO at a distance $L/2$ from the end of a semi-infinite waveguide. The setup is schematically shown in Fig. \ref{squeezingOPO}(a). The internal mode distribution in the waveguide can be modeled by a structured reservoir with a continuum of modes \cite{Hughes2007, Koshino2012, carmele2013, Bradford2013}. This model is particularly interesting for the case of short delays when the system is coupled at small distances from the end of the waveguide or from an additional cavity in the waveguide \cite{Hughes2007singleauthor} that acts as a mirror. However, for sufficiently long delays, losses within the waveguide, for instance in a coupled resonator optical waveguide structure \cite{Kamalakis2005}, the feedback strength might be substantially reduced. In that case a different setup (Fig. \ref{squeezingOPO}(b)) is necessary to achieve time-delayed feedback:

\paragraph*{Case (b), Feedback from an external mirror (Fig. \ref{squeezingOPO}(b))}

 This second possibility is to consider the DPO emitting directly into modes of an external mirror \cite{hetet2011, dornerzoller2002, eschner2001}, see Fig. \ref{squeezingOPO}(b). This setup is often used for semiclassical models of a laser in front of a mirror and is referred to as the Lang-Kobayashi model in laser physics \cite{Lang1980, Schulze2014}. In our case the description is fully quantum mechanical. Since the mirror can be placed at relatively long distances from the system, this model can particularly cover the range of long delays.


    
In both setups, the emitted signal interacts with the system again after a delay $\tau$, which is why we can model them on equal footing. In the appendix \refk{Equivalence of the proposed coherent feedback scheme to other approaches} we show their equivalence. Though, the difference is that for the first setup (Fig. \ref{squeezingOPO}(a)) there are two output fields accessible by measurement (at the waveguide and the mirror $M2$), in contrast to the second setup (Fig. \ref{squeezingOPO}(b)) where only the output field from mirror $M2$ is accessible.

Importantly, we consider the realistic case where dissipation is included, i.e., a fraction of the light emitted by the system through the mirror $M1$ couples with strength $\gamma_1$ to the structured feedback reservoir, which we introduce via the input-operator $b^1_{in}$ (cf. Fig. \ref{squeezingOPO}(a,b)). The remaining part of radiation which does not interact with the system again is modeled via another Markovian loss through a reservoir described with its input-operator $b^3_{in}$, the coupling strength of the system is expressed by $\gamma_3$. A similar approach considering an atom in front of a mirror is found in \cite{dornerzoller2002}. Moreover, the losses at the second cavity mirror $M2$ are described through the operator $b^2_{in}$ with coupling strength $\gamma_2$.

%

The system Hamiltonian of the pumped DPO is \cite{gardinerzoller2010book}
\begin{equation}
 H_{sys} = \hslash \omega_{0} c^{\dagger} c+\frac{i\hslash}{2}\,(\epsilon \mathrm{e}^{-i\omega_p t}  c^{\dagger 2}-\epsilon^{*} \mathrm{e}^{i\omega_p t} c^{2})
 \end{equation}
 
where
\begin{equation}
\epsilon = |\epsilon|\mathrm{e}^{i \beta}
\end{equation}
denotes the effective complex valued strength of the pump intensity and is proportional to the second order nonlinearity $\chi^{(2)}$. The operator $c$ is the photon annihilation operator of the cavity mode. The cavity mode frequency and the laser pump frequency are $\omega_0$ and $\omega_p$, respectively.

In the appendix section \ref{sect: Self-feedback as a quantum cascaded system} it is shown that from \refk{langevingleichung fuer rueckkopplung} the equation of motion of the operator $c$ subject to time-delayed feedback is 

  \begin{align}
 \dot{c}(t) =& -i \omega_0 c +\epsilon \mathrm{e}^{-i\omega_p t}c^{\dagger}(t) - G c(t) + \gamma_1 c(t-\tau) \notag \\
  &-\sqrt{\gamma_1}\,b^1_{in}(t)+\sqrt{\gamma_{1}}\,b^1_{in}(t-\tau) \notag \\& -\sqrt{\gamma_2}\,b^2_{in}(t)  -\sqrt{\gamma_3}\,b^3_{in}(t) \, ,
 \label{langevingleichung fuer rueckkopplung OPO}
 \end{align}
with $G=(2 \gamma_1 +\gamma_2 +\gamma_{3})/2$, and the input-operators $b^k_{in}(t)= 1/ \sqrt{2 \pi} \int_{-\infty}^\infty \! \mathrm{d}\omega  \, b^k_{0}(\omega) \mathrm{e}^{-i\omega t}$ which only depend on the initial reservoir-operator $ b^k_{t=0}(\omega)$ ($k=1,2,3$).  

We have set the phase-shift $\phi$ introduced by the mirror in the feedback loop in \refk{langevingleichung fuer rueckkopplung} to $\phi=\pi$, (cf. Fig. \refk{squeezingOPO}). We neglect here any frequency dependence of the phase-shift within the waveguide. Moreover we consider that no losses occur for the fields when reflected by the external mirror. According to the appendix section \refk{Equivalence of the proposed coherent feedback scheme to other approaches} this is modeled by choosing $\gamma_f=\gamma_1$ in equation \refk{langevingleichung fuer rueckkopplung}. The delay $\tau$ is the time needed by the field emitted at $M1$ to return back to $M1$, therefore $\tau= L / c_0$ with $L/2$ the system-mirror distance and $c_0$ the speed of light in the vacuum or in the waveguide.

We choose the pump frequency $\omega_p$ to be $2 \omega_0$ and move to a rotating frame of frequency $\omega_0$. In the rotating frame the equations become in matrix form


  \begin{align}
 \dot{\mathbf{c}}(t) =& \mathbf{A}\mathbf{c}(t) + \gamma_1 \mathbf{B}\mathbf{c}(t-\tau) \notag \\
  &-\sqrt{\gamma_1}\,\mathbf{b}^1_{in}(t) +\sqrt{\gamma_{1}}\, \mathbf{B}\mathbf{b}^1_{in}(t-\tau) \notag \\ & -\sqrt{\gamma_2}\,\mathbf{b}^2_{in}(t)-\sqrt{\gamma_3}\,\mathbf{b}^3_{in}(t) \, ,
 \label{langevin rot frame}
 \end{align}
 
where

\begin{align}
\mathbf{A}=\begin{pmatrix}
-G & \epsilon \\
\epsilon^{*} &-G  \\
\end{pmatrix} \quad \text{and} \quad
   \mathbf{B}= \begin{pmatrix}
\mathrm{e}^{i\omega_0 \tau} & 0 \\
0 & \mathrm{e}^{-i\omega_0 \tau}  \\
\end{pmatrix}
\end{align}

and using the operator vector notation,
\begin{equation}
\mathbf{O}(t)=\begin{pmatrix}
O(t)\\
O^{\dagger}(t)\\
\end{pmatrix} \, .
 \end{equation}

Since equation \refk{langevin rot frame} is linear we can solve it using Fourier transform techniques. We define the Fourier transformed field operators as 

\begin{equation}
\tilde{O}(w)= \dfrac{1}{\sqrt{2 \pi}} \int_{-\infty}^\infty \! \mathrm{d}t  \, \mathrm{e}^{-i\omega t} O(t) \, .
 \end{equation}

Note that

\begin{equation}
[\tilde{O}(\w)]^{\dagger}= [\dfrac{1}{\sqrt{2 \pi}} \int_{-\infty}^\infty \! \mathrm{d}t  \, \mathrm{e}^{-i\omega t} O(t)]^{\dagger}=\tilde{O}^{\dagger}(-\w)
 \end{equation}

so that $ \mathbf{\tilde{O}}(\w) = \begin{pmatrix}
 \tilde{O}(\w)\\
\tilde{O}^{\dagger}(-\w)\\
 \end{pmatrix}$.


The solution to equation \refk{langevin rot frame} reads:

 \begin{align}
 \mathbf{\tilde{c}}(w) = \mathbf{M}^{-1} \bigg [ &-\sqrt{\gamma_1} \begin{pmatrix} 
s(\w)&0\\
0&s^*(-\w)\\
 \end{pmatrix}\ \mathbf{\tilde{b}}^1_{in}(w) \notag \\& -\sqrt{\gamma_2} \mathbf{\tilde{b}}^2_{in}(\w) -\sqrt{\gamma_3} \mathbf{\tilde{b}}^3_{in}(\w) \bigg] \, ,
 \end{align}

where $s(\w)$ and $\mathbf{M}$ are given by:
\begin{equation}
s(\w)=1- \mathrm{e}^{-i(\omega-\omega_0) \tau} \, ,
\label{s}
 \end{equation}

\begin{align}
\mathbf{M}&=\begin{pmatrix}
i\omega+\frac{\gamma_2+\gamma_3}{2}+\gamma_1 s(\w)&-\epsilon\\
-\epsilon^{*}&i\omega+\frac{\gamma_2+\gamma_3}{2}+\gamma_1 s^*(-\w)
 \end{pmatrix} \, .
 \label{M}
 \end{align}

The function $s(\w)$ describes a frequency dependence of the loss through the channel described by $\gamma_1$. As it appears both in the matrix $\mathbf{M}$ and as the coupling strength for the field $b^1_{in}$, we can already expect important frequency dependent features in the squeezing spectra. We introduce $Z$ as an abbreviation for the matrix elements of $\mathbf{M}$: $\mathbf{M} \equiv \begin{pmatrix}
Z(\w)&-\epsilon\\
 -\epsilon^{*}&Z^*(-\w)\\
  \end{pmatrix} \, .$
After inverting $\mathbf{M}$ we find for the internal cavity mode


 \begin{align}
 \tilde{c}(\w) =& \mathcal{A}^{-}_{1}(\w)\tilde{b}^1_{in}(\w) +\mathcal{A}^{+}_{1}(\w)\tilde{b}^{1 \dagger}_{in}(-\w) \notag\\ + & \mathcal{A}^{-}_{2}(\w)\tilde{b}^2_{in}(\w) + \mathcal{A}^{+}_{2}(\w)\tilde{b}^{2 \dagger}_{in}(-\w)  \notag\\ + & \mathcal{A}^{-}_{3}(\w)\tilde{b}^3_{in}(\w) + \mathcal{A}^{+}_{3}(\w)\tilde{b}^{3 \dagger}_{in}(-\w) \, , \label{internal mode frequency}
 \end{align}

with
\begin{align}
 &\mathcal{A}^{-}_{1}(\w)=\frac{-\sqrt{\gamma_1} s(\w)Z^*(-\w)}{\Delta(\w)} && \quad \mathcal{A}^{+}_{1}(\w)=\frac{-\sqrt{\gamma_1} s^*(-\w)\epsilon}{\Delta(\w)} \notag\\
 &\mathcal{A}^{-}_{2}(\w)=\frac{-\sqrt{\gamma_2} Z^*(-\w)}{\Delta(\w)} && \quad \mathcal{A}^{+}_{2}(\w)=\frac{-\sqrt{\gamma_2}\epsilon}{\Delta(\w)} \notag\\
  &\mathcal{A}^{-}_{3}(\w)=\frac{-\sqrt{\gamma_3} Z^*(-\w)}{\Delta(\w)} && \quad \mathcal{A}^{+}_{3}(\w)=\frac{-\sqrt{\gamma_3}\epsilon}{\Delta(\w)} \, ,
 \end{align}

where

\begin{align}
\Delta(\w)=\text{det}(\mathbf{M})=Z(\w)Z^*(-\w)-|\epsilon|^2\, .
 \end{align}

Equation \refk{internal mode frequency} describes the internal field dynamics, provided the input fields are known. The next step is to calculate the output fields, since these determine the squeezing spectrum in an experiment \cite{Iida2012}.

\section{The output fields} 
So far we have determined the dynamics of the internal mode in terms of the input fields. The output is determined by the input-output relations \refk{input output relation} and \refk{input-output relation time bin bout} and rely on the input fields from mirror and waveguide, respectively. There are two detectable output fields, $b^1_{out}$ and $b^2_{out}$, accessible by measurement in the case covered by Fig. \ref{squeezingOPO}(a). However, when the feedback is realized by a mirror (Fig. \ref{squeezingOPO}(b)), the input- and output fields $b^1_{in/out}$ cannot be accessed by measurement in a straightforward way. They model the reservoir in which the ``in-loop'' excitation is lost after one round trip. In that case, only $b^2_{out}$ can be measured.

In the appendix (section \ref{Eliminating the in-loop field}) it is shown that in frequency space, the output fields take the general form


\begin{equation}
\tilde{b}^i_{out}(\w)= X_i(\w) \tilde{b}^i_{in}(\w)+Y_i(\w)\tilde{c}_{}(\w) \,.
\label{input-output relation fourier}
 \end{equation} 

where $X_i$ and $Y_i$ are defined in equation \refk{fourier input output anhang 3}.


For the direct input-output connection at $M2$ we simply have $X_2=1$ and $Y_2=\sqrt{\gamma_2}$ reducing the above expression to the standard input-output relation $\tilde{b}^2_{out}(\w)= \tilde{b}^2_{in}(\w)+\sqrt{\gamma_2}\tilde{c}_{}(\w)$, \citep{gardinerzoller2010book}. At the waveguide output equation \refk{input-output relation time bin bout} applies. We find after transformation in the rotating frame and in frequency space $X_1(\w)=-\mathrm{e}^{-i(\omega-\omega_0) \tau}$ and ${Y_1(\w)= \sqrt{\gamma_{1}}- \sqrt{\gamma_{1}}\mathrm{e}^{-i(\omega-\omega_0) \tau}=\sqrt{\gamma_1}s(\w)}$.

The relation \refk{input-output relation fourier} can be rewritten with \refk{internal mode frequency} as

\begin{align}
\tilde{b}^i_{out}(\w)= &\sum_{j} \mathcal{S}^-_{ij}(\w)\tilde{b}^j_{in}(\w) \\ 
+ &\sum_{j}\mathcal{S}^+_{ij}(\w) \tilde{b}^{j \dagger}_{in}(-\w) \, ,
\label{bout bin relation}
 \end{align}

where

\begin{align}
\mathcal{S}^-_{ij}(\w)& =Y_i(\w)\mathcal{A}^{-}_{j}(\w)+X_i(\w)\delta_{ij} \label{Sminus}\\ 
\mathcal{S}^+_{ij}(\w)&=  Y_i(\w)\mathcal{A}^{+}_{j}(\w) \, . 
\label{Splus}
 \end{align}

So far we have algebraically eliminated the in-loop field and the internal mode by re-expressing the output fields in terms of the input fields. That was straightforward since the equations are linear in frequency space. Similar approaches in linear quantum networks are found in \citep{gough2009}, \citep{gough2010}. Identifying $X_i$ and $Y_i$ from expression \refk{input-output relation fourier}, allows us to compute $\mathcal{S}^\mp_{ij}$.

 Given that the input fields are in the vacuum state, only antinormally ordered correlations are non-vanishing, that is

 \begin{align}
\ewqm{\tilde{b}^i_{in}(\w)\tilde{b}^{j\dagger}_{in}(-\w')}= \delta_{ij}\delta(\w+\w') \, ,
\label{anti normal correlation}
 \end{align}

all other correlations of the input fields are zero.

In this case we find for the reservoir output fields

 \begin{align}
\ewqm{\tilde{b}^i_{out}(\w)\tilde{b}^{i \dagger}_{out}(-\w')}= \mathcal{N}_{i}(\w)\delta(\w+\w') \notag\\
\ewqm{\tilde{b}^i_{out}(\w)\tilde{b}^i_{out}(\w')}= \mathcal{M}_{i}(\w)\delta(\w+\w') \, ,
\label{correlation output}
 \end{align}
 
 where 
 

\begin{align}
\mathcal{N}_{i}(\w)=& \sum_{j}  |\mathcal{S}^-_{ij}(\w) |^2\notag\\
\mathcal{M}_{i}(\w)= & \sum_{j}  \mathcal{S}^-_{ij}(\w)\mathcal{S}^+_{ij}(-\w) \, ,
\label{Mi and Ni}
\end{align}

where $\mathcal{S}^\pm_{ij}$ are defined in the equations (\ref{Sminus}-\ref{Splus}).

%

Furthermore we have the identity

\begin{align}
\sum_j |\mathcal{S}^-_{ij}(\w) |^2-|\mathcal{S}^+_{ij}(\w) |^2=1 \, .
\label{identity bogoliubov}
\end{align}

This ensures that the output fields satisfy the usual bosonic commutator relation. Considering the single input case (${b}^1_{in}=0$ or ${b}^2_{in}=0$) this relation simplifies to ${|\mathcal{S}^-_{ii}(\w) |^2-|\mathcal{S}^+_{ii}(\w) |^2=1}$. Then the output is simply a Bogoliubov transformation of the input \citep{gough2009, gough2010}. 

%
%

%
%
%
%

%
%
%
%

The system described by the equation \refk{langevin rot frame} is an open linear system. Energy is not conserved since there are loss and pump channels. We therefore need to analyze whether the system equilibrates at finite values, otherwise a static analysis for $t \rightarrow \infty$ is not reliable. To do so we carry out a stability analysis in the next section.

\section{Stability analysis} 
\label{Stability analysis}

Asymptotic stability requires that \textit{all} the roots $s$ of the characteristic equation of Eq. \refk{langevin rot frame}

\begin{equation}
\text{det}[s\mathbf{1}-\mathbf{A}-\gamma_1\mathbf{B}\mathrm{e}^{-\tau s}]=0 \, ,
\label{transcendental}
\end{equation}

lie in the complex open left half-plane \cite{thowsen1981, Shinozaki2007}. Since the transcendental equation \refk{transcendental} is difficult to solve, we use both a method using the Lambert-W function and an analytical treatment for delay-independent stability based on the Routh-Hurwitz criterion \cite{thowsen1981} to determine the asymptotic stability of our system.

We can dramatically reduce the complexity of the stability analysis by restricting the treatment to two important special cases where the coherent part of the field $b^1_{out,1}(t)$ and the time-delayed and phase shifted in-loop field $-b^1_{out,1}(t-\tau)$ are constructively or destructively interfering at the edge of mirror $M1$. 
We obtain constructive interference when the condition 
\begin{equation}
\w_0 \tau = (2n-1)  \pi 
\label{match 1}
\end{equation}
is matched and destructive interferences for
\begin{equation}
\w_0 \tau =2n  \pi \, , 
\label{match 2}
\end{equation}
where $n \in \mathbb{N}$ is any natural number.
In the following we restrict the stability analysis to these two cases.

We express the stability condition using a Lambert-W function approach. For the two special cases \refk{match 1} and \refk{match 2}, the characteristic equation \refk{transcendental} can be factorized into the form $(s-\alpha_1-\beta \mathrm{e}^{-\tau s})(s-\alpha_2-\beta \mathrm{e}^{-\tau s})=0$ with $\alpha_1=|\epsilon|-G$, $\alpha_2=-|\epsilon|-G$ and $\beta= \pm \gamma_1$ where ``$+$'' results from the condition $\w_0 \tau =2n  \pi $ and ``$-$'' from $\w_0 \tau = (2n-1)  \pi $. 

In this case the complex roots of Eq. (\ref{transcendental}) are given by

\begin{align}
s_i= \frac{1}{\tau}W_0(\tau \beta \mathrm{e}^{-\tau \alpha_i})+\alpha_i, \quad i=1, 2 \, ,
\end{align}
where $W_0$ is the principal branch of the Lambert W function \cite{Wright, Amann}.

It follows that the system described by Eq. (\ref{langevin rot frame}) is stable if and only if \cite{Shinozaki2007}

\begin{align}
& S^i_W(\alpha_i, \beta, \tau):=  
 Re[s_i]<0, \quad i=1, 2 \, .
\end{align}


We found that $ S^1_W \geq  S^2_W$, it is therefore sufficient to consider only $S^1_W$. Furthermore we can express $S^1_W$ in terms of dimensionless parameters as follows

\begin{align}
S^1_W=  
 \frac{1}{\tau} Re[W_0(\pm \gamma_1 \tau  \mathrm{e}^{\pm\gamma_1\tau( \tilde{\alpha}-1)})+( \tilde{\alpha}-1)] \, ,
\end{align} 

where again ``$+$'' or ``$-$'' result from the conditions (\ref{match 2}) and (\ref{match 1}), respectively, and

\begin{align}
 \tilde{\alpha}=\frac{|\epsilon|-(\gamma_2+\gamma_3)/2}{\gamma_1} \, .
 \label{tilde alpha}
\end{align}

This term represents the difference between the pump strength $|\epsilon|$ and losses $\gamma_2$ and $\gamma_3$ normalized by the feedback strength $\gamma_1$.

\begin{figure}
 \includegraphics[width=\linewidth]{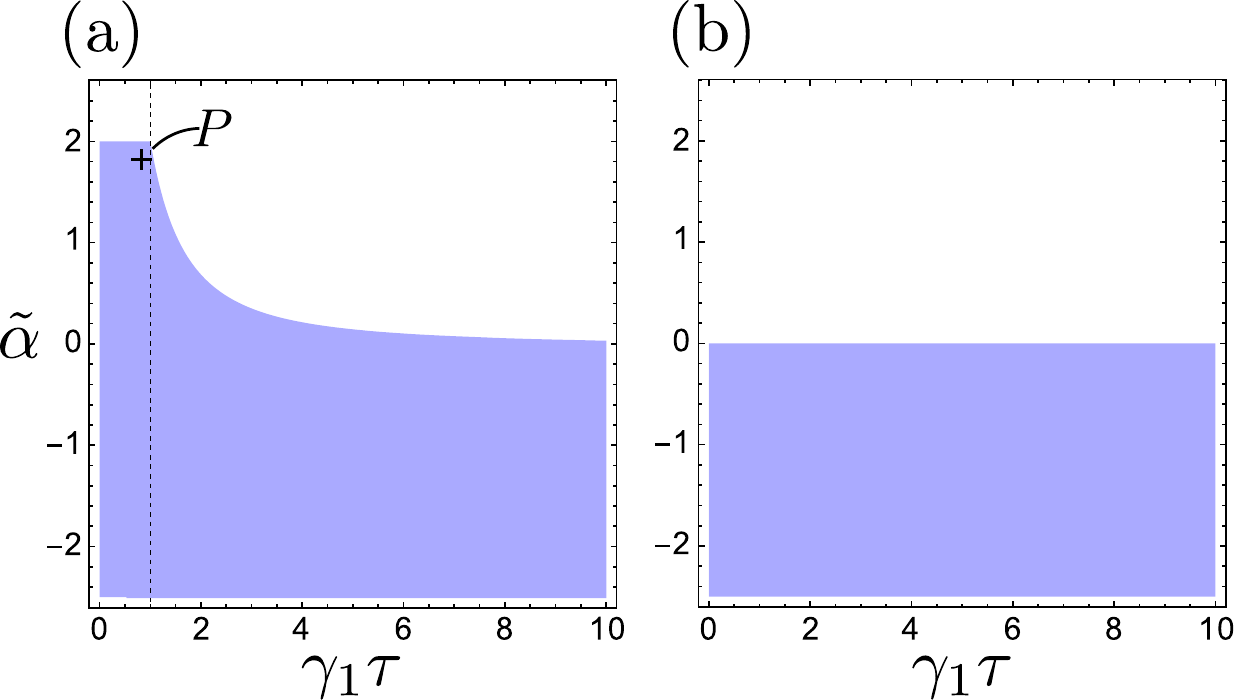}
 \caption{\label{fig: constructive lambert w} (Color online) Stability analysis in the case of constructive (a) and destructive (b) interference as a function of $\gamma_1 \tau$ and the effective pump strength $\tilde{\alpha}=(|\epsilon|-0.5(\gamma_2+\gamma_3))/\gamma_1 $, Eq. (\ref{tilde alpha}). The stable regions are the blue (shaded) ones, where $S^1_W<0$. (a) Constructive interference ($\w_0 \tau = (2n-1)  \pi $). The boundary line between the stable and unstable region is first constant with $\tilde{\alpha}=2$ for $\gamma_1 \tau <1$. At $\gamma_1 \tau=1$ there is a stability change. From that point, the stability boundary decreases with growing $\gamma_1 \tau>1$ toward $\tilde{\alpha} \rightarrow 0$. This means that for longer delays the stability becomes independent of the coupling $\gamma_1$ and of the delay time $\tau$. We call the region where $\gamma_1 \tau<1$ the \textit{short delay regime} and the region $\gamma_1 \tau>1$ the \textit{long delay regime}. The point $P$ represents a fixed value of ${|\epsilon|-(\gamma_{2}+\gamma_{3})/2}$ which is investigated in Fig. \ref{fig:spectrum3_S1_S2}. (b) Destructive interference ($\w_0 \tau = (2n-1)  \pi $). We observe that as long $\tilde{\alpha}<0$, i.e., $|\epsilon|<(\gamma_{2}+\gamma_{3})/2$ the system is stable. The stability is therefore independent of the coupling $\gamma_1$ to the feedback reservoir and of the delay time $\tau$.}
\end{figure}

We plot $S^1_W$ in dependence of the dimensionless parameters $\gamma_1 \tau$ and $\tilde{\alpha}$. The results for constructive and destructive interference are shown in Fig. \ref{fig: constructive lambert w}(a) and (b), respectively. 

\paragraph{Constructive interference, Fig. \ref{fig: constructive lambert w}(a):---} We find that the stability region is dependent on $\gamma_1 \tau$ and the parameter $\tilde{\alpha}$. For $\gamma_1 \tau<1$, the boundary line between unstable and stable region is constant with $\tilde{\alpha}=2$ corresponding to a pump strength $|\epsilon|=2 \gamma_1+(\gamma_2+\gamma_3)/2$. 

For $\gamma_1 \tau>1$ there is a stability change, the boundary line between unstable and stable region decreases slowly with $\gamma_1 \tau$. For sufficiently large $\gamma_1 \tau$ the boundary line approaches $\tilde{\alpha} \rightarrow 0$, i.e., $|\epsilon|\rightarrow (\gamma_2+\gamma_3)/2$. The stability then becomes independent of the coupling to the feedback reservoir and of the delay time $\tau$.

We use this boundary line to define what we call the \textit{short} and the \textit{long delay regime}:
We define the short delay regime for feedback times and strengths satisfying 
\begin{align}
\gamma_1 \tau <1 \, ,
\label{short feedback}
\end{align}

and furthermore we define the long delay regime for  
\begin{align}
\gamma_1 \tau >1 \, .
\label{long feedback}
\end{align}


\paragraph{destructive interference, Fig. \ref{fig: constructive lambert w}(b):---} In the case of destructive interference, we find that the system is stable for all $\gamma_1 \tau$ whenever $\tilde{\alpha}<0$, i.e.,  $|\epsilon|<(\gamma_2+\gamma_3)/2$. The stability is independent of the coupling to the feedback reservoir and of the delay time.

 Using a fully analytical method we can also give a lower limit where the system remains stable independently of the delay time $\tau$: %
from the Routh-Hurwitz criterion \cite{thowsen1981} it follows that the system \refk{langevingleichung fuer rueckkopplung OPO} is asymptotically stable over the whole range of possible delays $\tau$ if and only if the two following conditions are fulfilled:

\begin{enumerate}
\item The matrix $\mathbf{A}+\gamma_1\mathbf{B}$ is of Hurwitz type
\item The auxiliary equation \\ ${\text{det}[s\mathbf{1}-\mathbf{A}-\gamma_1\mathbf{B}(\frac{1-Ts}{1+Ts})^2]=0}$
has no roots on the imaginary axis for any $T\geq 0$.
\label{characteristic equation}
\end{enumerate}

%


For both cases we find analytically that stability is guaranteed
\emph{for all delays $\tau$} (as long as Eqs.~(\ref{match 1}) or ~(\ref{match 2})
hold) as long as
 \begin{equation}
\frac{\gamma_2+\gamma_3}{2}>|\epsilon| \,.
\label{stability criterion perfect match}
\end{equation}  
In case of destructive interference (Eq.~(\ref{match 2})) the system is
stable if and only if Eq.~(\ref{stability criterion perfect match}) is fulfilled. In contrast, for
constructive interference (Eq.~(\ref{match 1})) our analysis above (cf.
Fig.~\ref{fig: constructive lambert w}(a)) shows that there can be stable regions in which
Eq.~(\ref{stability criterion perfect match}) is not fulfilled. For long delays $\tau$, however,
Eq.~(\ref{stability criterion perfect match}) is recovered.

In summary, we find that the DPO losses dissipated into non-feedback reservoirs must be greater than the pump for the system to be stable. For constructive interference there exist additional losses ``created'' by the feedback which are particularly dominant in the short feedback range $
\gamma_1 \tau <1$ as depicted in Fig. \ref{fig: constructive lambert w} (a). For destructive interference only non-feedback terms contribute.


\section{Squeezing spectrum} 

We define observables called the output quadratures in Fourier space by \cite{gough2009}

\begin{equation}
\mathbf{X}^i_{out}(\w,\theta)=\mathrm{e}^{-i\theta} \mathbf{b}^i_{out}(\w)+\mathrm{e}^{i\theta} \mathbf{b}^{i \dagger}_{out}(-\w) \, ,
\label{quadrature}
 \end{equation}  

for fixed phases $\theta$. 
We calculate the variance of the quadratures at the output $i$ as 

\begin{equation}
\ewqm{\mathbf{X}^i_{out}(\w,\theta),\mathbf{X}^i_{out}(\w',\theta)}=\mathcal{P}_{i}(\w,\theta) \delta(\w+\w') \, ,
 \end{equation} 
where $\ewqm{a,b}=\ewqm{ab}-\ewqm{a}\ewqm{b}$ denotes the variance, where $\ewqm{\mathbf{X}^i_{out}(\w,\theta)}=0$ since the quadrature is simply a linear combination of the input fields which are in the vacuum-state.

The power spectral density $\mathcal{P}_{i}$ is referred to as the squeezing spectrum \cite{gough2009, Crisafulli2013} and takes the form  (cf. eq. \refk{correlation output})

\begin{align}
\mathcal{P}_{i}(\w,\theta)= &2\, Re[\mathrm{e}^{-i2\theta}\mathcal{M}_{i}(\w)]+\mathcal{N}_{i}(\w)+\mathcal{N}_{i}(-\w)-1 \, ,
\label{power spectral density}
 \end{align} 
 
with $\mathcal{M}_{i}$ and $\mathcal{N}_{i}$ as defined in Eq. (\ref{Mi and Ni}).

 
In particular for an (\textit{un}squeezed) coherent state or an (\textit{un}squeezed) vacuum state the variance in the quadratures, equation \refk{quadrature}, is equal to 1. This limit is referred to as the quantum noise limit. In contrast, a squeezed state is a non-classical state of light with the characteristic that for a certain phase $\theta$ the variance goes \emph{below} the quantum noise limit.

 
%
%
%

%
%
%
%


\section{Results} 

\begin{figure}
 \includegraphics[width=\linewidth]{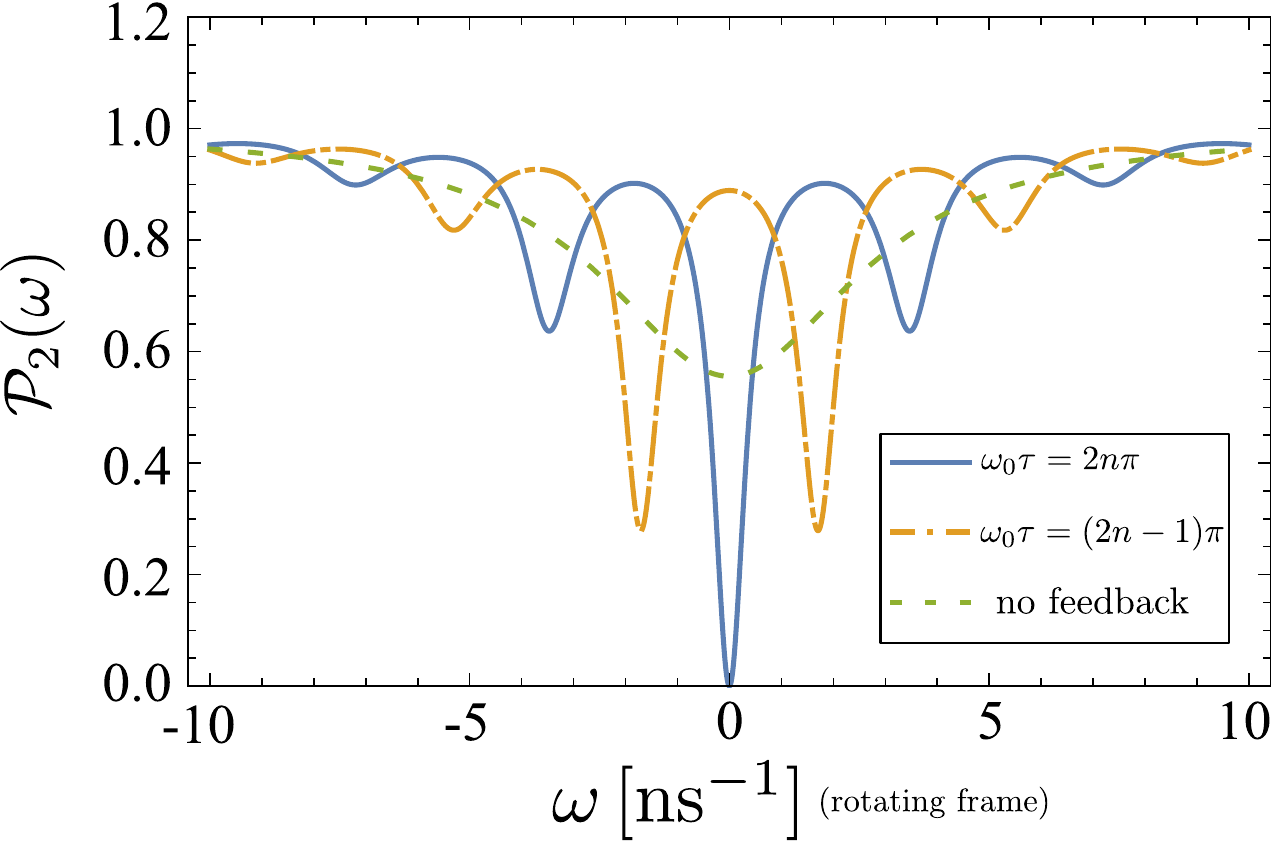}
 \caption{\label{fig:spectrum2_S2} (Color online) Long delay regime. Maximal squeezing spectra (in a frame rotating with $\w_0$) at the $M2$ output with and without feedback at threshold (in the feedback case) $|\epsilon|=\gamma_2 /2$ in the cases $\w_0 \tau =2n \pi $ and $\w_0 \tau = (2n-1)\pi$ for $\tau \neq 0$. The parameters are: $S=0.5$, $\gamma_1= \gamma_2=2 \, \text{ns}^{-1}$, $|\epsilon|=1 \, \text{ns}^{-1}$ and $\gamma_3=0$ (ideal feedback coupling). The spectrum is highly structured as a function of $\omega$ and $\tau$. For the case of destructive interference ($\w_0 \tau =2n \pi $) maximal squeezing is achievable at threshold at the central frequency $\omega_0$ ($\omega=0$ in the plots). On the other hand constructive interference cannot enhance the squeezing at $\w_0$. The side peaks show that the feedback scheme shifts the frequencies where squeezing occurs, however it is never maximal for the other frequencies than $\w_0$. The case without feedback corresponds to a double sided lossy cavity with respective loss rates $\gamma_1= \gamma_2=2 \, \text{ns}^{-1}$ pumped below threshold at $|\epsilon|=1 \, \text{ns}^{-1}$.}  

\end{figure}
\begin{figure}
 \includegraphics[width=\linewidth]{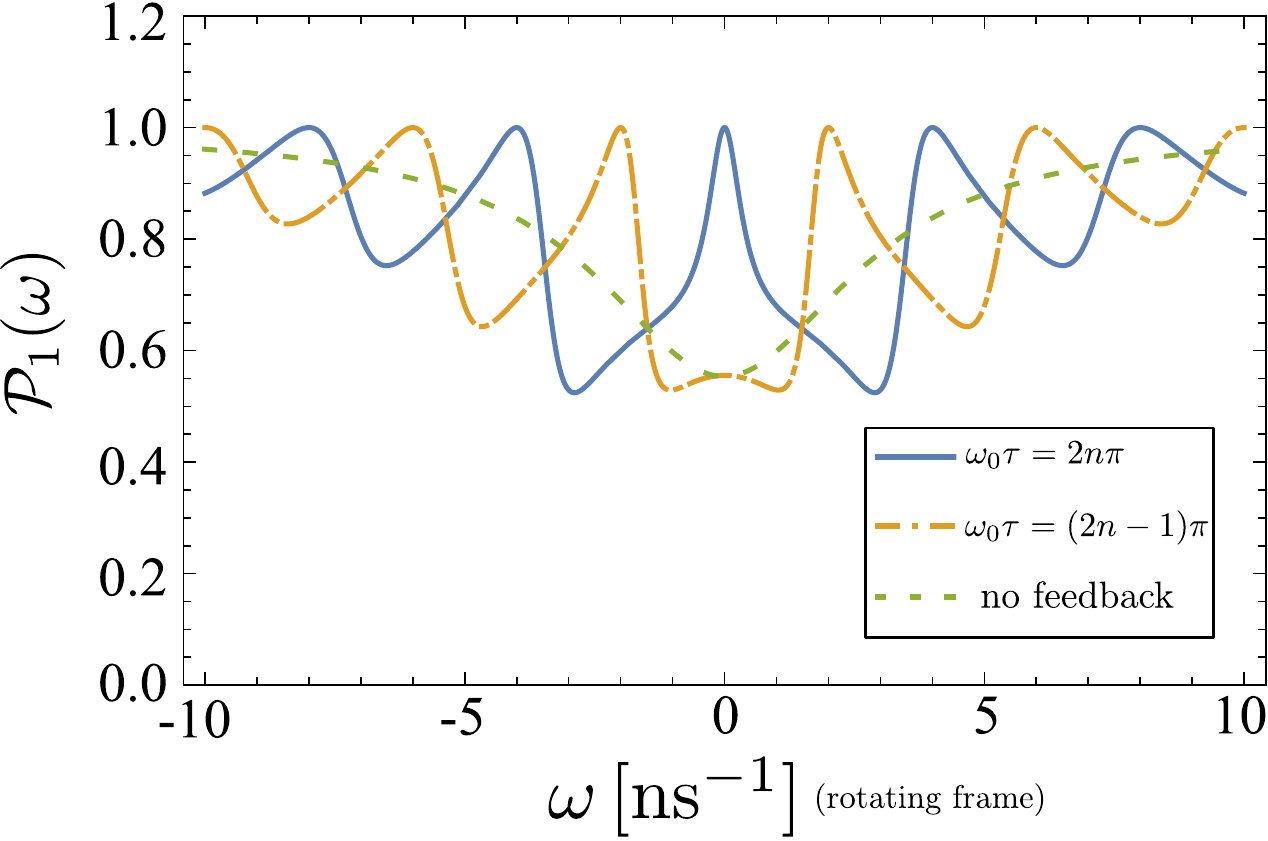}
 \caption{\label{fig:spectrum1_S1_long} (Color online) Long delay regime. Maximal squeezing spectra (in a frame rotating with $\w_0$) at the waveguide output with and without feedback at threshold $|\epsilon|=\gamma_2 /2$ in the cases $\w_0 \tau =2n \pi $ and $\w_0 \tau = (2n-1)\pi$ for $\tau \neq 0$. The parameters are: $S=0.5$, $\gamma_1= \gamma_2=2 \, \text{ns}^{-1}$, $|\epsilon|=1 \, \text{ns}^{-1}$ and $\gamma_3=0$ (ideal feedback coupling). The spectrum is highly structured as a function of $\omega$ and $\tau$. For the case of destructive interference ($\w_0 \tau =2n \pi $) only noise remains as an output at threshold at the central frequency $\omega_0$ ($\omega=0$ in the plots), thus ${\mathcal{P}_1(\omega_0)=1}$ corresponding to the quantum noise limit and no squeezing is achieved. On the other hand for constructive interference we observe squeezing of the output field at $\omega_0$. However for long delays $\tau$ in both cases $\gamma_{2}/2> |\epsilon|$ is required for stability, which decreases squeezing below $50\%$. For comparison, the green dashed line marks the squeezing spectrum for the case in which the feedback channel is replaced by a Markovian reservoir with loss rates $\gamma_1= \gamma_2=2 \, \text{ns}^{-1}$ pumped below threshold at $|\epsilon|=1 \, \text{ns}^{-1}$.}  
\end{figure}

\begin{figure*}
 \includegraphics[width=\linewidth]{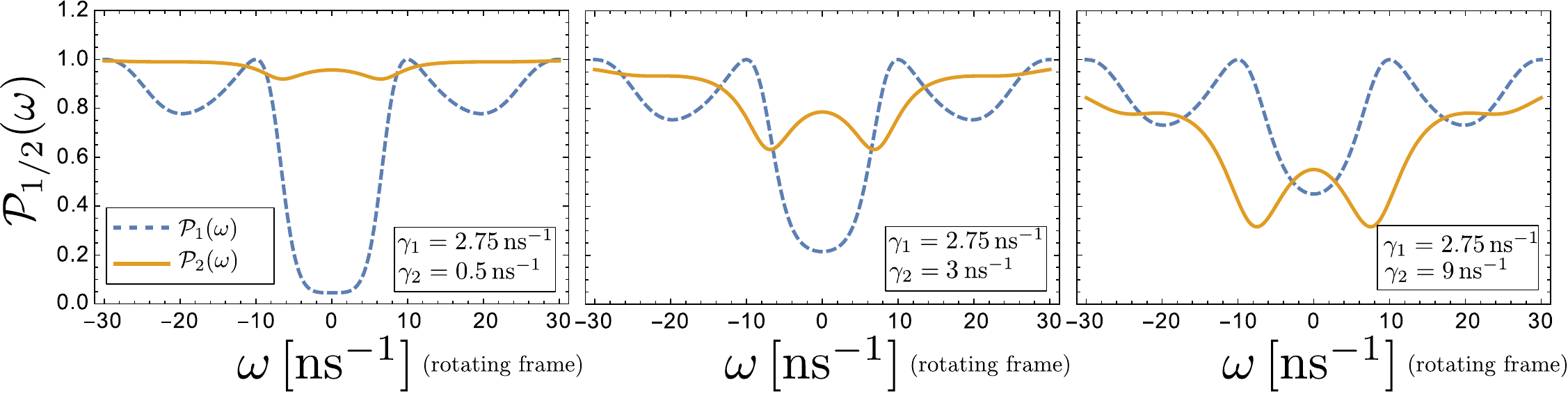}
 \caption{\label{fig:spectrum3_S1_S2} (Color online) Short delay regime. Comparison of the squeezing spectra $\mathcal{P}_1$ and $\mathcal{P}_2$ (in a frame rotating with $\w_0$) in the case of constructive interference, i.e., $\w_0 \tau = (2n-1)\pi$ for $\tau \neq 0$. The parameters are: $S=0.1 $, $\gamma_1= 2.75 \, \text{ns}^{-1}$,$\gamma_2 \in \{0.5, 3, 9 \} \, \text{ns}^{-1}$, and a fixed loss-pump strength $|\epsilon|-\gamma_{2}/2=5\, \text{ns}^{-1}$. The situation corresponds to the stable point denoted by $P_1$ in Fig. \ref{fig: constructive lambert w}(a). We see that while $\gamma_2$ increases, the squeezing in the output of the waveguide is diminished, whereas the squeezing at the mirror $M2$ output is enhanced. Note that for the present parameters the uncontrolled case ($\gamma_1=0$) is not stable and therefore cannot be shown (stability would require $|\epsilon|<\gamma_2/2$).}
\end{figure*}

To understand the delay dependence of the spectrum it is instructive to look at the expression \refk{s} and the matrix \refk{M}. The term $s(\omega)$ represents a frequency dependence of the loss through the channel described by $\gamma_1$. Compared to the uncontrolled scenario, it will cause a frequency-dependent self-energy of the spectrum. We expect that the Lorentzian form of the spectrum for the uncontrolled system becomes multi-peaked depending of the phase of the in loop field. Since the influence of $s(\omega)$ scales with the feedback strength $\gamma_1$, the frequency dependent effects increase with growing $\gamma_1$. We will proceed to analyze this behavior for different parameters.

For the numerical evaluation we choose the parameters $2\theta=\beta+\pi$, $\w_0=1 \, \text{fs}^{-1}$ and 

\begin{equation}
\tau= (S+10^{-6} \delta)\pi  \, \text{ns}.
\label{tau divided}
\end{equation}

Here, we divide $\tau$ into two contributions: First, the scaling parameter $S$ determines the length of the feedback time in $\pi$ ns and is chosen such that $S\omega_0 \, \text{ns}$ is a natural number. Second, the tuning parameter $\delta \in \lbrace 0,1 \rbrace$ is such that either the condition $\w_0 \tau = (2n-1)  \pi $ (destructive interference) or $\w_0 \tau =2n \pi $ (constructive interference) is fulfilled. We do focus on these two regimes of destructive and constructive interference to study the highest and lowest possible squeezing performance.

In the following the scaling parameter $S$ is within $\{0.1,\, 0.5\} $ so that the system-mirror distances (see Fig. \ref{squeezingOPO}) are in the range of few centimeters. 


The other parameters $|\epsilon|$ (pump), $\gamma_1$ (feedback coupling) and $\gamma_{2/3}$ (loss) are varied to address the experimental feasibility of the following numerical results. 
In the following plots
the quantum noise limit is equal to $1$. Whenever the values of the spectrum get below the quantum noise limit we have a squeezed quadrature. Perfect squeezing, i.e., maximal noise reduction, is obtained when the variance in the quadrature is zero.

To obtain a detailed analysis of the spectrum at the two output channels {${b}^1_{out}$} and {${b}^2_{out}$} we will investigate them for two different parameter sets, representing short and long delay times, defined in the section \ref{Stability analysis} Eqs. (\ref{short feedback}-\ref{long feedback}) as $\gamma_1 \tau>1$ and $\gamma_1 \tau<1$, respectively. For each output, we will start with the analysis of long delay times, where we will find strongly frequency-modulated squeezing spectra. Second we will investigate the short delay regime for parameters at which time-delayed feedback is crucial for the stabilization of the system. 


\subsection{Output spectrum at \boldmath${b}^2_{out}$}
In this section, we discuss the squeezing spectrum for the output channel located at mirror $M2$, i.e., ${b}^2_{out}$.

\textit{Long delay regime:---} First, we consider ideal feedback ($\gamma_3=0$) in the long delay regime. Non-ideal feedback ($\gamma_3 \neq 0$) is discussed further below. Because of the parameter set in this case, we are able to discuss both constructive interference, i.e., condition \refk{match 1}, and destructive interference, i.e., condition \refk{match 2}, at mirror $M1$. The stability analysis, section (\ref{Stability analysis}), shows that the system is stable for $|\epsilon|<\gamma_2/2$ over the whole range of $\tau$-values, if equations \refk{match 1} or \refk{match 2} are fulfilled. The used parameters are ${S=0.5}$ and $\gamma_1= \gamma_2=2 \, \text{ns}^{-1}$. We evaluate the squeezing spectrum $\mathcal{P}_2$ at the threshold value $|\epsilon|\rightarrow \gamma_2/2$. The spectrum is shown in Fig. \ref{fig:spectrum2_S2}. For comparison, we also plot the spectrum for a system in which the feedback decay channel is replaced by a Markovian reservoir (dashed line). In contrast to the free evolving spectrum without feedback, we observe that the spectrum is highly structured as a function of $\omega$. The frequency modulation is on the order of $O(\pi/\tau)$, therefore we observe that for longer delays the number of peaks increases within the frequency bandwidth. In particular, we find a strong enhancement of squeezing for specific frequencies ${\omega}$ compared to the uncontrolled case. Most importantly, for the case of destructive interference ($\w_0 \tau =2n \pi $) maximal squeezing ($\mathcal{P}_2=0$) is attained at threshold at the central frequency $\omega=\omega_0$ ($\omega=0$ in the rotating frame in Fig. \ref{fig:spectrum2_S2}). Here, because of the destructive interference, the feedback channel is ``closed'' so that effectively the system behaves like a single ended cavity \footnote{Although our scheme involves a double-sided cavity, for destructive interference it shows the squeezing performance of a single-ended cavity pumped at threshold without feedback \citep{collettgardiner1984}.}. The case of constructive interference ($\w_0 \tau =(2n-1) \pi $) shows a good squeezing performance for the first side peaks, whereas no noise reduction is observed at $\omega_0$. 

\textit{Short delay regime:---} Let us now reduce the delay time such that $\gamma_1\tau<1$. We will analyze the system for parameters where time-delayed feedback with \emph{constructive} interference (condition \refk{match 1}) is crucial for the stability of the system. In particular, we will use the parameters marked as ``$P$'' in Fig. \ref{fig: constructive lambert w}(a). We set the delay scaling parameter $S$ (Eq. \refk{tau divided}), to be $0.1$ and analyze the squeezing properties for a set of fixed loss-pump differences $|\epsilon|-\gamma_{2}/2=5\, \text{ns}^{-1}$, in which we vary $\gamma_2 \in \{0.5,3,9 \} \, \text{ns}^{-1}$ and choose the feedback coupling strength to be ${\gamma_1=2.75\, \text{ns}^{-1}}$, which corresponds to $\gamma_1\tau\simeq0.864$. Again we put $\gamma_3=0$ for simplicity. We can only discuss the case for constructive interference, since destructive interference as well as no feedback at all would make the system unstable. 

The calculated spectra (short delay, constructive interference) are shown in Fig. \ref{fig:spectrum3_S1_S2} (solid line). We observe again that the spectrum $\mathcal{P}_2$ is structured as a function of $\omega$, with two main side peaks. We find that while $\gamma_2$ increases the squeezing at the mirror $M2$ output is enhanced, the qualitative behavior does not change, however it can only be maximal ($\mathcal{P}_2\rightarrow 0$) in the limit situation $\gamma_2 \rightarrow \infty$ (not shown). In this case the feedback mechanism can be neglected compared to the losses occurring at the mirror $M2$ output. Below we will analyze the waveguide output spectrum $b^ 1_{out}$ for the same parameters to get a direct comparison with $b^2_{out}$.   

For both cases, i.e., long and short feedback times, we have also investigated the case ${\gamma_3 \neq 0}$ (not shown). The only effect of ${\gamma_3 \neq 0}$ is to reduce the squeezing in the spectra, while it does not change the qualitative frequency behavior.

\subsection{Output spectrum at \boldmath${b}^1_{out}$}
Next we address the question, whether the output spectrum depend on the channel of observation, $b^1_{out}$ or $b^2_{out}$, cp. Fig. \ref{squeezingOPO}. For this we discuss the waveguide signal, Fig. \ref{squeezingOPO}(a).

\textit{Long delay regime---} For the discussion of the long delay situation we evaluate the squeezing spectrum of the ${b}^1_{out}$ for the same set of parameters as in the long delay analysis of the ${b}^2_{out}$ output spectrum. The spectrum is shown in Fig. \ref{fig:spectrum1_S1_long}. In the case of long delays for both cases of destructive and constructive interference, $\gamma_{2}/2> |\epsilon|$ is required for stability, which decreases squeezing below $50\%$. 
In case of destructive interference we observe $\mathcal{P}_1=1$ at the central frequency $\omega=\omega_0$ ($\omega=0$ in the rotating frame). In this case only vacuum noise remains at the output of the waveguide, since all the radiation from the cavity is cancelled out by its time-delayed counterpart. For comparison, we again also plot the squeezing spectrum \emph{without} feedback, i.e., when replacing the feedback channel by a Markovian reservoir. 

\textit{Short delay regime---} 
 We first discuss the spectrum at the output of the waveguide in the short feedback regime in the case of constructive interference, i.e., condition \refk{match 1} for the same set of parameters as in the short delay analysis of the ${b}^2_{out}$ output spectrum. The spectra are shown in Fig. \ref{fig:spectrum3_S1_S2} (dashed line). Similar to the $b^2_{out}$ spectrum, the spectra are highly structured as a function of $\omega$ with one main peak at the central frequency $\omega=\omega_0$ ($\omega=0$ in the rotating frame) and two side peaks. We find that for small $\gamma_2$ values, i.e., small losses at the mirror $M2$ the squeezing in the ${b}^1_{out}$ output at the central frequency $\omega_0$ is very high ($\mathcal{P}_1\rightarrow 0$ for $\gamma_2 \rightarrow 0$). For increasing $\gamma_2$ the squeezing in the output of the waveguide ($M1$) is reduced, whereas the squeezing at the mirror $M2$ output is enhanced. In this case there are two independent inputs which cannot interfere to cancel the fluctuations and increase the squeezing in the quadrature.   

As outlined in the stability analysis, additional losses are generated by the feedback in the case of constructive interference, thereby the system remains stable. The spectrum is broadened around $\w_0$ so that the squeezing is maximal over a large frequency region. In the lossless case ($\gamma_{2}=0$, not shown here) we find perfect squeezing at the threshold curve (boundary between stable and unstable region). Furthermore we observe a frequency region of constant squeezing in the spectrum for stable points in the short delay regime as $P$ (c.f. Fig. \ref{fig: constructive lambert w}(a)). On the other hand, when passing the border between the short and long delay regime, the spectra have a dip at $\w_0$ (not shown) which grows with the value of the feedback coupling strength $\gamma_1$.

\section{Conclusion}

We have shown how quantum coherent time-delayed feedback of Pyragas type \cite{pyragas1992} can be used to control and enhance the squeezing performance at the output of a cavity containing a degenerate parametric oscillator. We proposed two physically different setups to introduce the feedback that could both be modeled on equal footing. Our main result is that coherent feedback causes a frequency dependent modification of the power spectral density of the output field quadratures, compared to the uncontrolled case. For the application of control, the Lorentzian form of the uncontrolled squeezing spectrum becomes sharply multi-peaked due to frequency dependent interferences within the feedback loop. In particular, we find a strong enhancement of squeezing for specific delay times. A thorough stability analysis shows that in the case of constructively interfering signals within the feedback loop, new loss channels are created, meaning that the pump intensity of the laser can be increased while the system remains stable. This becomes particularly important in the short delay regime as this allows to enhance the squeezing in one of the systems output channels far below the quantum noise limit.

\begin{acknowledgments}
We would like to thank Alexander Carmele for very helpful discussions. 
We acknowledge support from Deutsche Forschungsgemeinschaft (DFG) through SFB 910 ``Control of self-organizing nonlinear systems'' (project B1 and project A1).
\end{acknowledgments}

\appendix

\section{Description of quantum-coherent time-delayed feedback}
\label{appendix A}

\begin{figure}[h!]
 \includegraphics[width=\linewidth]{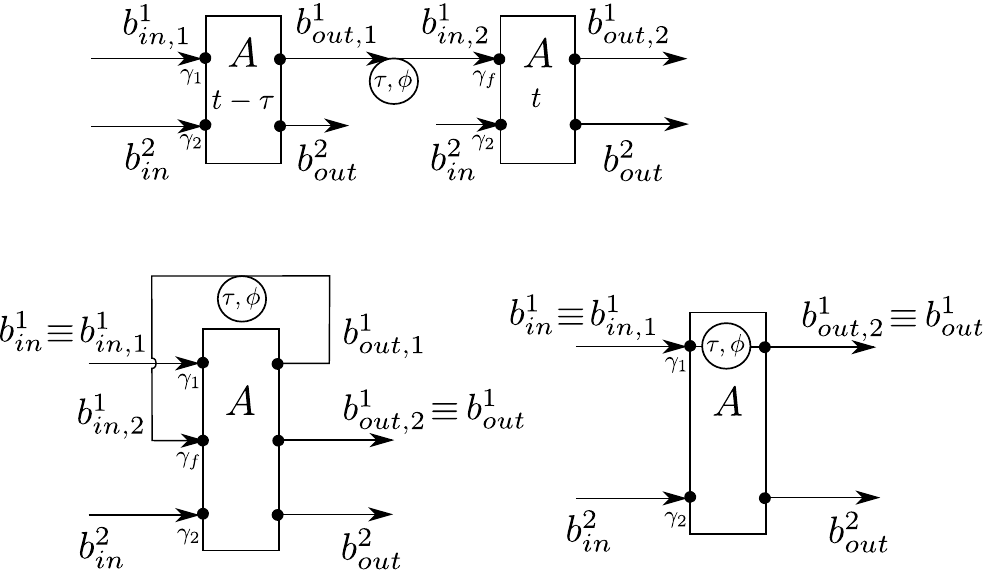}
 \caption{\label{fig:model} (Color online) Our feedback scheme. Top: A system is coherently driven in a cascaded fashion by a past version of itself via the feedback in-loop field $b^1_{in,2}(t)=\mathrm{e}^{i\phi} b^1_{out,1}(t-\tau)$. The additional input $b^2_{in}$ takes account of additional drives or losses of the system independent of  $b^1_{in}$. {Left bottom:} Same physical setup represented by a loop. {Right bottom:} The in-loop field can be eliminated, its only purpose is to modify the internal dynamic. Note that $b^1_{out}$ never influences the system again.}
\end{figure}

In this appendix we present a method to describe quantum-coherent time-delayed feedback using the input-output theory \citep{gardinercollett1985inputoutput, gardinerzoller2010book}. Starting from the classical description of quantum cascaded systems, we will show how the formalism can be used to describe quantum coherent feedback, where previous versions of the system drive the present one. Note that the input-output theory uses a Markov approximation that is only valid for a weak system-reservoir coupling. Therefore it is important to ensure that the feedback setup does not alter the coupling of the system to a continuum of reservoir modes.    
A principal difficulty lies in the fact that no master equation exists since the system becomes highly non-Markovian in the presence of feedback. An attempt following the precedent idea to solve the non-linear system by embedding the feedback system in a larger space is found in \cite{grimsmo2015}. Also Pyragas type feedback control schemes can be realized where the system is driven towards a desired steady-state \cite{grimsmo2014}.

\subsection{Coherent feedback as a quantum cascaded system}
\label{sect: Self-feedback as a quantum cascaded system}

In what follows we want to investigate the problem of a system driven coherently by a past version of itself. We shall use the input-output theory applied to cascaded systems, \citep{gardinercollett1985inputoutput}, \citep{gardinerzoller2010book}.  

To begin let us consider the classical case of a separable quantum system with total Hamiltonian $H_{sys}$ that can be decomposed into two subsystems $A$ and $B$ with Hamiltonian $H_{A}$ and $H_{B}$ respectively. The total system is interacting with the modes of an external reservoir with Hamiltonian $H^1_{R}  =  \hslash \int_{-\infty}^\infty \! \mathrm{d}\omega  \, \omega  b^{\dagger}(\omega) b^{}(\omega)$, where $b^{(\dagger)}(\omega)$ are bosonic field modes satisfying $\ko{b^{}(\omega)}{b^{\dagger}(\omega')}=\delta(\w-\w')$. The systems $A$ and $B$ interact with the reservoir through their respective operators $c_A$ and $c_B$.

In terms of the input-output theory, we now drive the input of the first system $A$ with the bosonic field induced by the external reservoir modes. The interaction-Hamiltonian of $A$ with the reservoir is therefore (within the rotating wave approximation)
  
\begin{align}
H_{Int,1}  =  i \hslash \sqrt{\frac{\gamma_1}{2 \pi}} \int_{-\infty}^\infty \! \mathrm{d}\omega  \,  [ b^{\dagger}(\omega)c_A-\text{h.c.}] \,.
\label{ww1 system a}
\end{align}
 
Furthermore we assume that the second system $B$ is driven by the time-delayed and phase-shifted output field of the first one so that the interaction Hamiltonian takes the form:
 
\begin{align}
H_{Int,2}  = i \hslash \sqrt{\frac{\gamma_{f}}{2 \pi}} \int_{-\infty}^\infty \! \mathrm{d}\omega  \,  [ b^{\dagger}(\omega)  \mathrm{e}^{i\omega \tau} \mathrm{e}^{i\phi} c_B-\text{h.c.}]
\label{ww2 system b}
\end{align}


Where $\gamma_1$ and $\gamma_{f}$ are the coupling elements which are constant. This is the result of a standard approximation often called first Markov-approximation \cite{gardinercollett1985inputoutput} which holds true when all coupling from the system to the reservoir is within a narrow bandwidth of frequency \cite{gardinerzoller2010book}. The term $\omega \tau$ is such that $\tau c_0$ is the distance between the two systems, with $c_0$ the speed of light. Furthermore we consider an additional phase-shift $\phi$ of the field which may be induced by an external boundary condition, e.g., by a mirror. The "$f$" in the coupling constant $\gamma_f$ stands for feedback and points out that the system $B$ is coupled to the time-delayed light-field emitted by $A$. We extend the treatment by additionally coupling $A$ to another reservoirs $H^{k}_R$, $k \in \{2,3,..  \} $ independent of $H^1_{B}$ in the form \refk{ww1 system a}. We distinguish the reservoirs via the superscript in $H^{k}_R$. The Fig. \ref{fig:model} shows schematically the cascaded system. From the standard input-output theory we derive the following Langevin-equation for an arbitrary operator $X$ of the systems $A$ or $B$ \cite{gardiner1993cascade}

\begin{widetext}
 \begin{subequations}
 \label{langevin sys}
\begin{align}
\dot{X} = & -\frac{i}{\hbar} \ko{X}{H_{A}+H_{B}} \\
&  -\sum_{k} \, \bigg\{ \ko{X}{c_A^\dagger}\left(\frac{\gamma_k}{2}c_A+\sqrt{\gamma_k}b^k_{in}(t)\right) -\left(\frac{\gamma_k}{2}c^{\dagger}_A+\sqrt{\gamma_k}b^{k \dagger}_{in}(t)\right)\ko{X}{c_A} \bigg\} \label{langevin sys1}\\
& - \ko{X}{c_B^\dagger}\left(\frac{\gamma_{f}}{2}c_B+\sqrt{\gamma_1\gamma_{f}}c_A(t-\tau)\,\mathrm{e}^{i\phi} +\sqrt{\gamma_f}b^1_{in}(t-\tau)\,\mathrm{e}^{i\phi} \right) \notag \\
&+ \left(\frac{\gamma_{f}}{2}c^{\dagger}_B+\sqrt{\gamma_1\gamma_{f}}c^{\dagger}_A(t-\tau)\,\mathrm{e}^{-i\phi} +\sqrt{\gamma_f}b^{1 \dagger}_{in}(t-\tau)\,\mathrm{e}^{-i\phi} \right)\ko{X}{c_B}  \, , \label{langevin sys2} 
\end{align} 
\end{subequations}
\end{widetext}


We have defined the input operator $b^k_{in}(t)= 1/ \sqrt{2 \pi} \int_{-\infty}^\infty \! \mathrm{d}\omega  \, b^k_{0}(\omega) \mathrm{e}^{-i\omega t}$ which only depends on the initial reservoir-operator $ b^k_{t=0}(\omega)$. 

We see from equation \refk{langevin sys} that only system $A$ has an effect on system $B$ whereas $B$ does not influence $A$. Thus we have an unidirectional coupling field from $A$ to $B$. 

This setup can now be modified to represent time-delayed feedback: 
In the following we consider only one additional reservoir $H_R^2$ beside $H_R^1$. To describe a system driven coherently by a past version of itself through a feedback channel, we now go one step further and let the two subsystems $H_A$ and $H_B$ become the same system $H_{sys}$, which means that $c_A=c_B=c$. We consider the equation of motion for the operator $c$, which from \refk{langevin sys} takes the form 

  \begin{align}
\dot{c} =& -\frac{i}{\hslash}\ko{c}{H_{sys}} - \frac{\gamma_1 +\gamma_2+\gamma_{f}}{2} c(t)- \sqrt{\gamma_1 \gamma_{f}} \mathrm{e}^{i \phi} c(t-\tau) \notag\\
&- \sqrt{\gamma_1}b^1_{in}(t) - \sqrt{\gamma_2}b^2_{in}(t) - \sqrt{\gamma_{f}} \mathrm{e}^{i \phi} b^1_{in}(t-\tau) \, ,
\label{langevingleichung fuer rueckkopplung}
\end{align}


Equation \refk{langevingleichung fuer rueckkopplung} is a time-delayed differential equation. This is a quite general equation, indeed no assumptions have yet been made about the system Hamiltonian nor the initial statistics of the reservoir. The input-operators $b_{in}$ only depend on the initial condition $b_{0}$. They can therefore be interpreted as noise terms if the system and reservoir are initially factorized and $b_{0}$ is in a incoherent state. The delayed term $b^1_{in}(t-\tau)$ indicates that past fluctuations have an influence on the dynamics of the present system. We now simply have the situation of a system interacting with an input field giving rise to an output field which interacts with the system again after a delay $\tau$ and a phase shift $\phi$. Moreover causality has to be preserved, that means that the system at later times should not influence the previous system. This is naturally ensured in the present case, where $sys_1$ (previous) influence $sys_2$ (later) but the reverse direction is forbidden. Therefore the input-output theory for quantum cascade systems is well suited for the treatment of a quantum coherent description of autonomous time-delayed feedback as it allows input to output connections with preservation of causality. A graphical representation of the cascade is given in Fig. \ref{fig:model} top.

\subsection{Description of feedback in terms of loops} 

We define output fields related by the standard input-output relation \cite{gardinerzoller2010book}

\begin{equation}
b^k_{out,\alpha}(t)= b^k_{in,\alpha}(t)+\sqrt{\gamma_{k,\alpha}}\, c(t) \, .
\label{input output relation}
\end{equation}

The subscript $\alpha$ labels the input- and output-ports which are driven by fields from the same reservoir $k$. The problem is schematically shown in Fig. \ref{fig:model} bottom left. We omit the subscript in the case where the operator $b^k_{in/out}$ only acts on one port of the system.

We remark that \refk{langevingleichung fuer rueckkopplung} can be directly obtained by coupling the system to two bosonic fields $b^1_{in,1}(t)$ and $b^1_{in,2}(t)$ with respective strength $\gamma_1$ and $\gamma_{f}$ by letting

\begin{equation}
 b^1_{in,2}(t)=\mathrm{e}^{i \phi} b^1_{out,1}(t-\tau) \, .
 \label{input output in loop}
\end{equation}
Using the input-output relation \refk{input output relation} we can directly connect the in-loop input to the free input  $b^1_{in}$, that is

\begin{equation}
 b^1_{in,2}(t)=\mathrm{e}^{i \phi} ( b^1_{in,1}(t-\tau)+\sqrt{\gamma_1}c(t-\tau)) \, .
 \label{input output in loop 2}
\end{equation}

Inserting the new input $b^1_{in,2}$ in \refk{langevin sys} without \refk{langevin sys2} reproduce \refk{langevingleichung fuer rueckkopplung}. 

This is an interesting point since it gives an intuitive picture of autonomous feedback. In fact the field  $b^1_{in}$ induced by the reservoir gives rise to the in-loop field $b^1_{out,1}$. This is propagating back into the system again to become the input $b^1_{in,2}$ after having experienced a phase-shift $\phi$ and a delay-time $\tau$. The only purpose of this in-loop field is to modify the internal dynamics of the system and may then be lost in form of the output $b^1_{out,2}$. That means that the information of the past system is only stored for the duration of one loop in the feedback-channel and may then be forgotten. Therefore only one $\tau$ appears in the equation.  


In the present case we do not allow $b^1_{out,2}$ to influence the system again, but we can straightforwardly extend equation \refk{langevingleichung fuer rueckkopplung} to include this case. However, the present treatment is sufficient for a number of problems dealing with time-delayed feedback in open quantum systems when the coupling between system and reservoir is weak, see, e.g., \cite{carmele2013, grimsmo2014, dornerzoller2002}. We will discuss that in the next section.

\subsection{Equivalence of the proposed coherent feedback scheme to other approaches}
\label{Equivalence of the proposed coherent feedback scheme to other approaches}

We reconsider now the Hamiltonians of section \refk{sect: Self-feedback as a quantum cascaded system}. We will show that our coherent feedback-scheme is equivalent to other approaches found in the literature.

Consider the case where the interactions $H_{int,1}$ and $H_{int,2}$ of the system with the reservoir $H^1_R$ are done with the same operator $c_A=c_B=c$ and are of same strength, that is $\gamma_1=\gamma_f$. Furthermore consider a phase-shift $\phi=\pi$. 

We introduce the interaction-Hamiltonian $H_{int}$ as the sum of $H_{int,1}$ and $H_{int,2}$.

\begin{align}
H_{int} &= H_{int,1}+H_{int,2} \notag \\
&= 2 \hslash \sqrt{\frac{\gamma_1}{2 \pi}} \int_{-\infty}^\infty \! \mathrm{d}\omega  \,  [\sin(\w \tau/2) B^{\dagger}(\omega)c+\text{h.c.}] \,.
 \label{interaction hamiltonian equivalence}
\end{align}

Here we have defined the new bosonic operator

\begin{align}
B(\w)= \mathrm{e}^{-i \w \tau/2} b(\w) \,.
\end{align}

This is exactly the interaction Hamiltonian found when coupling a system weakly to an external continuum of modes shaped by a mirror, as it is done, e.g., in \cite{dornerzoller2002}, \cite{carmele2013} and \cite{hein2014}.

Since $H^1_{R}  =  \hslash \int \! \mathrm{d}\omega  \, \omega  b^{\dagger}(\omega) b^{}(\omega) =  \hslash \int \! \mathrm{d}\omega  \, \omega  B^{\dagger}(\omega) B^{}(\omega)$, the system dynamics described by the total Hamiltonian $H=H_{sys}+H^1_R+H_{int}$ is therefore totally equivalent in the two approaches. The input-output theory as already outlined before particularly gives a more intuitive picture of the feedback problem.

\subsection{Eliminating the in-loop field} 
\label{Eliminating the in-loop field}

 It should be mentioned that it is impossible to measure the in-loop field without destroying the quantum coherence of the feedback loop, so that the only purpose of this field is to modify the internal dynamics. It is suitable to derive a direct relation between the free output $b^1_{out,2} \equiv b^1_{out}$ to the free input $b^1_{in}$ since it avoids dealing with non standard commutation relation as \refk{kommutator input 2} when considering, e.g., correlations of the inputs. This is easily done using the input-output relation. We have
 \begin{equation}
 {b}^1_{out}(t)= \mathrm{e}^{i \phi}{b}^1_{in}(t-\tau) +\mathrm{e}^{i \phi}\sqrt{\gamma_{1}}{c}_{}(t-\tau) +\sqrt{\gamma_{f}}{c}_{}(t)
 \label{input-output relation time bin bout}
  \end{equation}   
A graphical representation is given by Fig. \ref{fig:model} right bottom. In the main part of the paper a Fourier representation of equation \refk{input-output relation time bin bout} is needed. Using the Fourier transformed field operators defined by
\begin{equation}
\tilde{O}(w)= \dfrac{1}{\sqrt{2 \pi}} \int_{-\infty}^\infty \! \mathrm{d}t  \, \mathrm{e}^{-i\omega t} O(t) \, ,
 \end{equation} 
we can transform \refk{input-output relation time bin bout} into

\begin{equation}
\tilde{b}^1_{out}(\w)=  \mathrm{e}^{i \phi}  \mathrm{e}^{-i \omega \tau} \tilde{b}^1_{in}(\w)+(\sqrt{\gamma_{1}} \mathrm{e}^{i \phi}\mathrm{e}^{-i \omega \tau}+ \sqrt{\gamma_{f}}) \tilde{c}_{}(\w) \,.
\label{fourier input output anhang 1}
 \end{equation}
 
 In the same way, the input-output relation for ${b}^2_{in}/{b}^2_{out}$, ${b}^2_{out}(t)={b}^2_{in}+\sqrt{\gamma_2}c(t)$, is simply 
 \begin{equation}
 {\tilde{b}^2_{out}(\w)=  \tilde{b}^2_{in}(\w)+ \sqrt{\gamma_{2}} \tilde{c}_{}(\w)} \,.
 \label{fourier input output anhang 2}
 \end{equation}
 
In a general form, covering both relations we can write the equations \refk{fourier input output anhang 1} and \refk{fourier input output anhang 2} as 
  
\begin{equation}
\tilde{b}^i_{out}(\w)= X_i(\w) \tilde{b}^i_{in}(\w)+Y_i(\w)\tilde{c}_{}(\w) \,,
 \label{fourier input output anhang 3}
 \end{equation}

where $X_i$ and $Y_i$ are defined in the equations \refk{fourier input output anhang 1}, \refk{fourier input output anhang 2} and are used as an abbreviation in the main part of the paper.

\subsection{Commutator relations} 

The input fields $b^i_{in}$ obey the bosonic commutator relation

\begin{equation}
 \ko{b^i_{in}(t)}{b^{j \dagger}_{in}(t')}=\delta_{ij} \delta(t-t')\,,
\label{kommutator input}
\end{equation}

and in case of a vacuum input the only non-vanishing correlation is

\begin{equation}
\ewqm{b^i_{in}(t)b^{i \dagger}_{in}(t')}=\delta(t-t')\,.
\label{correlation input}
\end{equation}

For the in-loop input $b^1_{in,2}$ this relation are modified since $b^1_{in,2}$ is not independent of $b^1_{in}$ and the system operator $c$. We have for the commutator 

\begin{align}
 \ko{b^1_{in,2}(t)}{b^{1 \dagger}_{in,2}(t')}= \delta(t-t') \quad \text{for} \quad |t-t'|< \tau \,.
\label{kommutator input 2}
\end{align}

which can be found using the equations \refk{input output relation}, \refk{input output in loop} and for reasons of causality: $\ko{b^{1 (\dagger)}_{out,1}(t')}{c^{(\dagger)}(t)}=0$ for $t-t'<\tau$, (see, e.g., \cite{gardinerzoller2010book}).

Thus the in-loop field obeys the canonical commutator relation only for $|t-t'|<\tau$ but not for time differences greater than the loop time.
 Ensuring that the time difference $|t-t'|$ is less than the delay $\tau$ guarantees that the two-time commutator remains canonical as it evaluates between fields in coexistence in the loop.

\subsection{Pyragas type quantum control scheme} 
If we consider the situation where the phase-shift is $\phi = \pi$ and the coupling is $\gamma_1=\gamma_{f}=\gamma$ (perfect coupling) and $\gamma_2=0$ (lossless) the feedback channel \textit{engineers} a quantum control scheme which is of the Pyragas type \cite{pyragas1992}.  In this scheme equation \refk{langevingleichung fuer rueckkopplung} becomes

\begin{align}
\dot{c} =& -\frac{i}{\hslash}\ko{c}{H_{sys}} -\gamma_1 [c(t)-  c(t-\tau)] \notag\\
&- \sqrt{\gamma_1}b^1_{in}(t) + \sqrt{\gamma_{1}}  b^1_{in}(t-\tau) \, .
\end{align}

 In this scheme the term $\gamma_1[c(t)-c(t-\tau)]$ acts as a control force and vanishes in case of stabilization, when the system reaches a steady-state or is $\tau$-periodic, i.e., if ${c(t)=c(t-\tau)}$. The output field is then using \refk{input-output relation time bin bout} ${b^1_{out}(t) = - b^1_{in}(t-\tau) + \sqrt{\gamma_1} [c(t)-c(t-\tau)]}$. When the system reaches a steady-state we will have $c(t)=c(t-\tau)$ and in consequence the system-operator will be stabilized. For the output field we then have $b^1_{out}(t) = - b^1_{in}(t-\tau)$. That means only noise remains as an output in case of stabilization. The Pyragas scheme is non-invasive since the control-force vanishes in case of stabilization. Pyragas-type stabilization schemes have been used in coherent feedback control, e.g., to reach fast steady-state convergence in open quantum systems \cite{grimsmo2014}. We presented recently how to use a Pyragas scheme to control entanglement in a quantum node network \cite{Hein2015}.

\bibliography{squeezingfinal}

\end{document}